\documentclass[twocolumn]{aastex701}

\usepackage[utf8]{inputenc}

\usepackage{makecell}
\usepackage{tabularx}

\usepackage[starfontserif]{starfont}
\usepackage{amsmath}
\DeclareSymbolFont{starfontsym}{OT1}{sts}{m}{n}
\DeclareMathSymbol{\mathJupiter}{\mathord}{starfontsym}{106}

\usepackage{graphicx}
\usepackage{hyperref}

\begin{document}

\title{A constraint on the density of Jupiter's moon Thebe from primordial dynamics}
\author[0000-0002-2478-3084]{Ian R. Brunton}
\affiliation{Division of Geological and Planetary Sciences, California Institute of Technology}
\email[show]{\href{mailto:brunton@caltech.edu}{brunton@caltech.edu}}

\author[0000-0002-7094-7908]{Konstantin Batygin}
\affiliation{Division of Geological and Planetary Sciences, California Institute of Technology}
\email{kbatygin@caltech.edu}

\begin{abstract}

Of the 97 known satellites in the Jovian system, the individual masses and densities of each moon have only been determined for six of them: the four Galileans, Amalthea, and Himalia. In this letter, we derive a prediction for the mean density (and mass) of Thebe, Jupiter's sixth largest regular moon, obtaining a lower limit of $\rho_\text{T}\gtrsim1.0$ g/cm$^3$ ($m_\text{T}\gtrsim 5\times 10^{20}$ g). In particular, this value emerges as a key constraint within the context of the resonant transport model for the origins of Jupiter's interior satellites. Expanding on this theory, here we carry out simulations of the simultaneous gravitational shepherding of Amalthea and Thebe via the resonant influence of inward-migrating Io during Jupiter's disk-bearing epoch. We find that owing to overstability of resonant dynamics facilitated by the circumjovian disk's aerodynamic drag, Thebe's smaller radius (compared to that of Amalthea's) requires a higher density to ensure its terminal orbital distance exceeds that of Amalthea's, as it does today. With multiple current and upcoming space missions devoted to \textit{in situ} exploration of the Jovian system, a proper measurement of Thebe's mass provides an avenue towards empirical falsification or confirmation of our theoretical model for the dynamical evolution of Jupiter's inner moons. 
\\
\end{abstract}

\section{Introduction}
\label{sec:intro}

Establishing a detailed theory for the formation of giant planets marks a critical milestone in painting a full picture of the Solar System's origins. Equally important, unraveling the genesis of Jupiter's natural satellites provides an essential step towards this broader understanding. While numerous theories of satellite formation have been proposed over the past half-century (e.g. \citealt{Lunine&Stevenson82,CanupWard02,CanupWard06,Mosquiera&Estrada03a,Mosquiera&Estrada03b}), efforts have intensified over the last decade (e.g. \citealt{Sasaki+10,Miguel&Ida16,Cilibrasi+18,Cilibrasi21,Shibaike+19,RonnetJohansen20,BatyginMorbidelli20,Madeira+21}). Collectively, these recent models share an emerging consensus: Jupiter's Galilean satellites likely assembled within a circumplanetary disk at relatively large orbital radii before migrating inward via satellite-disk interactions (see review by \citealt{McKinnon23} and references therein). Within this evolving framework, however, a fundamental question remains unanswered: how do Jupiter's \textit{minor} regular moons -- Metis, Adrastea, Amalthea, and Thebe -- which occupy tightly bound orbits \textit{interior} to the Galileans, fit into this developing paradigm? In a recent paper, \citet{Brunton&Batygin25}, we began examining this question, focusing first on the largest of these inner satellites, Amalthea.  

The key idea underpinning our hypothesis is that Amalthea formed at a substantial orbital distance from Jupiter alongside the Galilean satellites. As the neighboring Io accumulated enough mass to trigger inward type-I migration, Amalthea became captured into an interior mean-motion resonance with the larger moon, subsequently being shepherded inward as Io continued its journey toward Jupiter. Ultimately, Amalthea's inward transport stalled as it entered Jupiter's magnetospheric cavity, settling at its present-day orbit. Because Amalthea's mass and density are well-determined \citep{Anderson+05}, numerical modeling of this resonant transport scenario enabled us to place meaningful constraints on the thermodynamic properties of the circumjovian nebula. Specifically, we demonstrated that stable resonant migration capable of reproducing Amalthea's present orbital configuration \textit{requires} that the circumplanetary disk possessed an aspect ratio in excess of $h/r \gtrsim$ 0.08. 

Now we continue our investigation into the origin of Jupiter’s inner satellites by turning the focus to the second largest of the inner moons, Thebe, which currently revolves at $a\sim 3.11\ R_\text{J}$, \textit{between} the orbits of Io and Amalthea. Thebe's limited dataset indicates that, like Amalthea, it formed in a distant region of the circumjovian nebula \citep{Takato+04}, and in our model, its modern position would so too indicate it would have been a part of the resonant transport induced by Io's migration. But because Thebe’s mass (and thus density) is unknown, little could be gleaned from it regarding its transport and a link to primordial nebular properties that is not already found by examining the more tightly constrained Io-Amalthea dynamics. Most excitingly, however, the unknown mass of Thebe opens a separate door for confirmation or falsification of our theoretical model. 

What ultimately matters in determining whether Io’s long-range transport can account for the orbital architecture of \textit{both} Amalthea and Thebe, is (1) the magnitude of overstability in the resonant transport, and (2) the relative timescale of decay between that of Amalthea and Thebe that is induced by aerodynamic drag in the disk. The first criterion is what we investigated in \citet{Brunton&Batygin25}, and is most appreciably governed by the above stated value in $h/r$. As we will show below, introducing Thebe into the picture does not alter this result.\footnote{Because of its small size, Thebe also succumbs to overstability once locked in Io's interior resonance. However, Thebe's smaller radius from that of Amalthea entails a somewhat \textit{dampened} vigor of  overstable librations \textit{up to} a density of $\rho_\text{T} \sim 1.47$ g/cm$^3$. This is evident from the expressions we derived in \citet{Brunton&Batygin25} for the equilibrium eccentricity, $e_0$ (eq. 22-24), and the growth rate for the amplitude of libration, $\langle\dot{A}\rangle$ (eq. 28), which in the illustrative limits we discussed in that work, evolve with satellite properties as $e_0\propto (R\rho)^{1/3}$ and $\langle\dot{A}\rangle\propto R\rho$, respectively.}

The second criterion, however, is strongly correlated to the intrinsic properties of Amalthea and Thebe themselves. In particular, during the final stages of inward migration -- when the satellites approach the magnetospheric cavity -- the sub-Keplerian headwind in the circumjovian nebula can induce orbital decay that is comparable to (or even shorter than) that dictated by resonant shepherding. Thus, the relative inward migration rates of Amalthea and Thebe at this critical stage depend sensitively on their own respective density and size. Because aerodynamic drag scales as $\dot{a}_{\mathrm{gas}} \propto R\,\rho^{-1}$ \citep{Adachi+76}, the key unknown property in determining the relative decay timescales is Thebe’s density, $\rho_{\text{T}}$. If Thebe is significantly \textit{under-dense} (i.e., similar to the under-dense Amalthea, $\rho_{\text{A}}$), its smaller radius might cause it to migrate inward excessively quickly, potentially violating its currently observed exterior orbit relative to Amalthea. Consequently, quantifying the conditions under which the current orbital ordering can emerge imposes meaningful constraints on Thebe’s unknown density, $\rho_{\text{T}}$, motivating the numerical investigation presented in this work.

The remainder of this Letter is structured as follows. In \S \ref{sect:NumExp}, we present a suite of numerical experiments involving Io, Amalthea, and Thebe, designed to reproduce and clarify the terminal orbital architecture of these satellites. We then summarize our key results in \S \ref{sect:Discussion}, discussing their broader implications, and highlighting caveats alongside observational avenues toward confirming or falsifying the theoretical model outlined in our work.

\section{Numerical Experiments}
\label{sect:NumExp}

\begin{figure*}[t]
    \includegraphics[width=\textwidth]{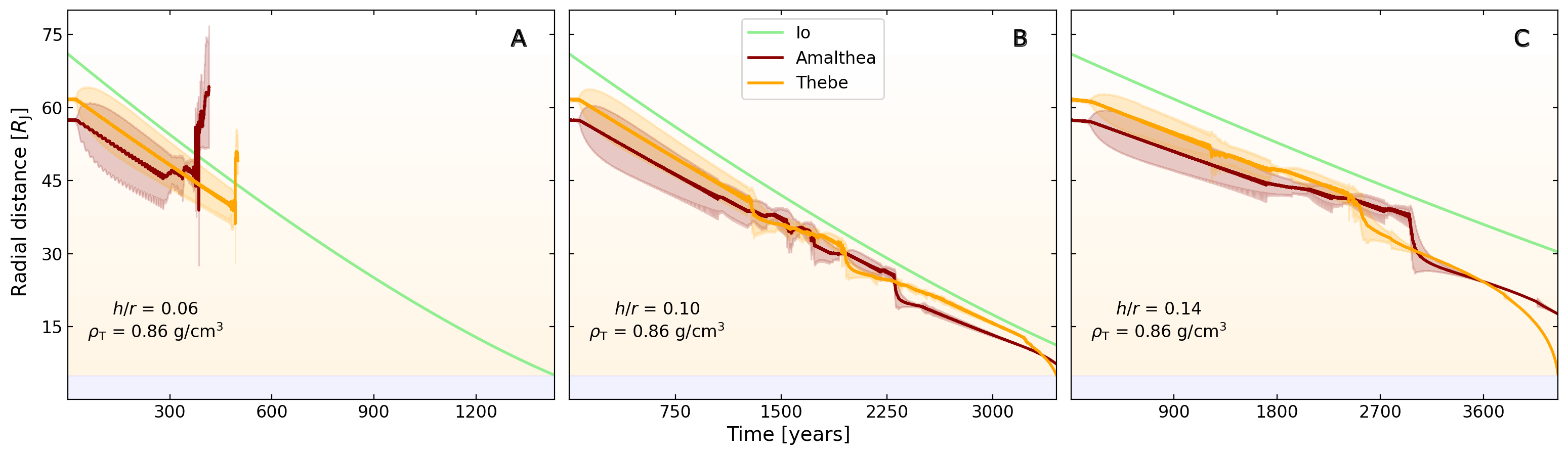}
    \caption{Three simulations of the scenario described in the text, and more thoroughly described in \citet{Brunton&Batygin25}. Io, Thebe, and Amalthea, form in a zone of satellite building blocks near $r_0\sim0.1\ R_\text{H}$, where Io subsequently captures the smaller moons on interior mean motion resonant orbits once it type-I migrates inward. Each plot shows the joviancentric distance evolution of the three moons. Shaded envelopes around Amalthea and Thebe trace the difference in their perijove and apojove, thus characterizing their eccentricity in resonance. The light blue band marks the magnetospheric cavity beginning at the disk's truncation radius, $R_\text{t}=5\ R_\text{J}$. In each run, Thebe's density is given an equivalent value to Amalthea's known density, $\rho_{T} = \rho_\text{A} = 0.86$ g/cm$^3$, and the disk aspect ratio is varied from $h/r=0.06$, $0.10$, and $0.14$.}
    \label{fig:TopTheMigPanels}
\end{figure*}

\begin{figure*}
    \includegraphics[width=\textwidth]{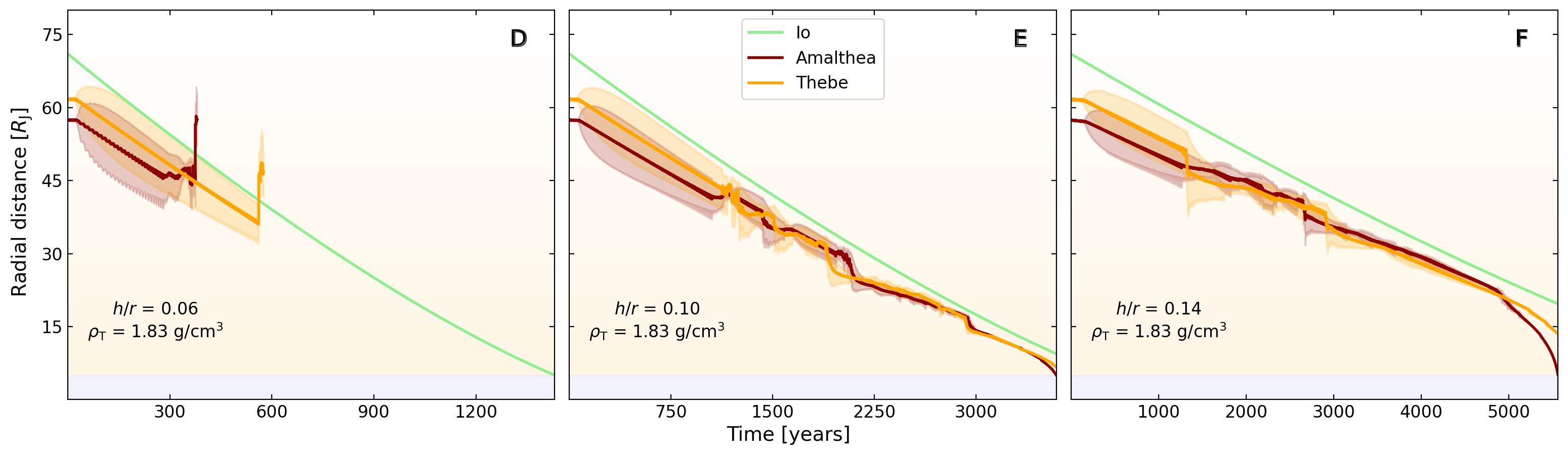}
    \caption{Three simulations with the same setup as Figure \ref{fig:TopTheMigPanels}, except Thebe is now given a Callisto-like density, $\rho_{T} = \rho_\text{C} = 1.83$ g/cm$^3$.}
    \label{fig:BotTheMigPanels}
\end{figure*}

In addition to the fact that Thebe is currently in between Io and Amalthea, \citet{Takato+04} suggested that its surface spectra points to it having formed in the distant regions of the disk. Working then with the model we constructed in \citet{Brunton&Batygin25} to explain Amalthea's path to the inner Jovian system, we now consider Io's simultaneous resonant transport of Thebe and Amalthea from the outer regions of the disk, inward to their modern orbital neighborhood. 

\subsection{Setup}

To investigate this scenario, we simulate the dynamical evolution of the satellites amidst a viscous circumjovian disk, utilizing the \texttt{Rebound}/\texttt{Reboundx} code with the integrator, \texttt{mercurius} \citep{rebound,reboundx,reboundmercurius}, with the effects of the disk implemented via the same methods and equations delineated in \citet{Brunton&Batygin25}. We set a constant disk aspect ratio between $h/r=0.08$--0.15 (the key constraint that emerged in our previous work), and vary Thebe's density between realistic values of $\rho_{\text{T}} = 0.5$--3.0 g/cm$^3$. The disk profile is fixed to a reference surface density of $\Sigma_0=4000$ g/cm$^2$ at $r_0=0.1\ R_\text{H}$ with profile index, $s=1.25$.\footnote{As shown in our previous work, the details of resonant transport are relatively insensitive to changes in $\Sigma$ or $s$.} Io is initialized on a circular orbit at $a_\text{I} = r_0$, while Thebe and Amalthea are input randomly between $a_{\text{T,A}} = 0.75$--$0.95\ a_\text{I}$.\footnote{Both moons experience chaotic dynamics as they undergo resonant transport, and thus their starting location has little affect on which moon concludes their transport at the interior orbit.} 

We then evolve the system until the following events occur: (1) Amalthea or Thebe reaches the truncation radius of the disk at $R_\text{t}= 5\ R_\text{J}$, (2) Amalthea or Thebe is scattered outward beyond Io and unable to recircularize interior to, or (3) a collision between any of the three bodies.\footnote{While the mutual gravitational interactions between Amalthea and Thebe are included in our calculations, they are negligible given the small masses of the two tiny moons ($m \lesssim 10^{-9}\ M_\text{J}$). Moreover, their dynamical evolution very rarely results in an Amalthea-Thebe collision ($\ll$ 1\%), with the two instead undergoing high-eccentricity orbit crossings as seen in the exemplary runs of Figures \ref{fig:TopTheMigPanels} \& \ref{fig:BotTheMigPanels}.} The outcome is then only labeled a ``success" if Amalthea reaches the truncation radius \textit{before} Thebe. 

\subsection{Simulations}

Let us begin by examining a set of illustrative examples of these simulations. As a first approximation, consider a scenario in which Thebe has an Amalthea-like density, $\rho_\text{T}=\rho_\text{A}=0.86$ g/cm$^3$.
Panels A, B, and C of Figure \ref{fig:TopTheMigPanels} depict three such exemplary runs with the disk aspect ratio varied as $h/r = 0.06$, 0.10, and 0.14. These runs can be compared to Figure 3 of \citet{Brunton&Batygin25} to see that the qualitative relationship between overstability and $h/r$ that we investigated in that work is maintained with the introduction of Thebe. The key detail, however, is that now in the two simulations shown in which the small bodies can be delivered by Io to the inner Jovian system (panels B and C), Thebe ultimately ends up on an \textit{interior} orbit to that of Amalthea. The determinant time period for this ordering is during the final phase of Amalthea and Thebe's evolution into the magnetospheric cavity, in which we see the aerodynamic drag timescale outpace that of Io's type-I migration, $\tau_a|_\text{gas} \lesssim \tau_a|_\text{typI}$, and Thebe is more strongly affected by this owing to its smaller radius than the larger Amalthea. Clearly, Thebe's aerodynamic timescale must be enhanced relative to this baseline expectation, and the only avenue to achieve this is to consider a higher density for the unknown $\rho_\text{T}$. 

Figure \ref{fig:BotTheMigPanels}, with panels D, E, and F, now shows three exemplary runs in which we set Thebe's density equivalent to that of Callisto's, $\rho_\text{T}=\rho_\text{C}=1.83$ g/cm$^3$. Again overstability is evident in each run, but most notably, in the two simulations in which Amalthea and Thebe are delivered to the inner Jovian system (E and F), Amalthea enters the magnetospheric cavity interior to Thebe, and the modern-day ordering is replicated.

\begin{figure}
    \centering
    \includegraphics[width=1\linewidth]{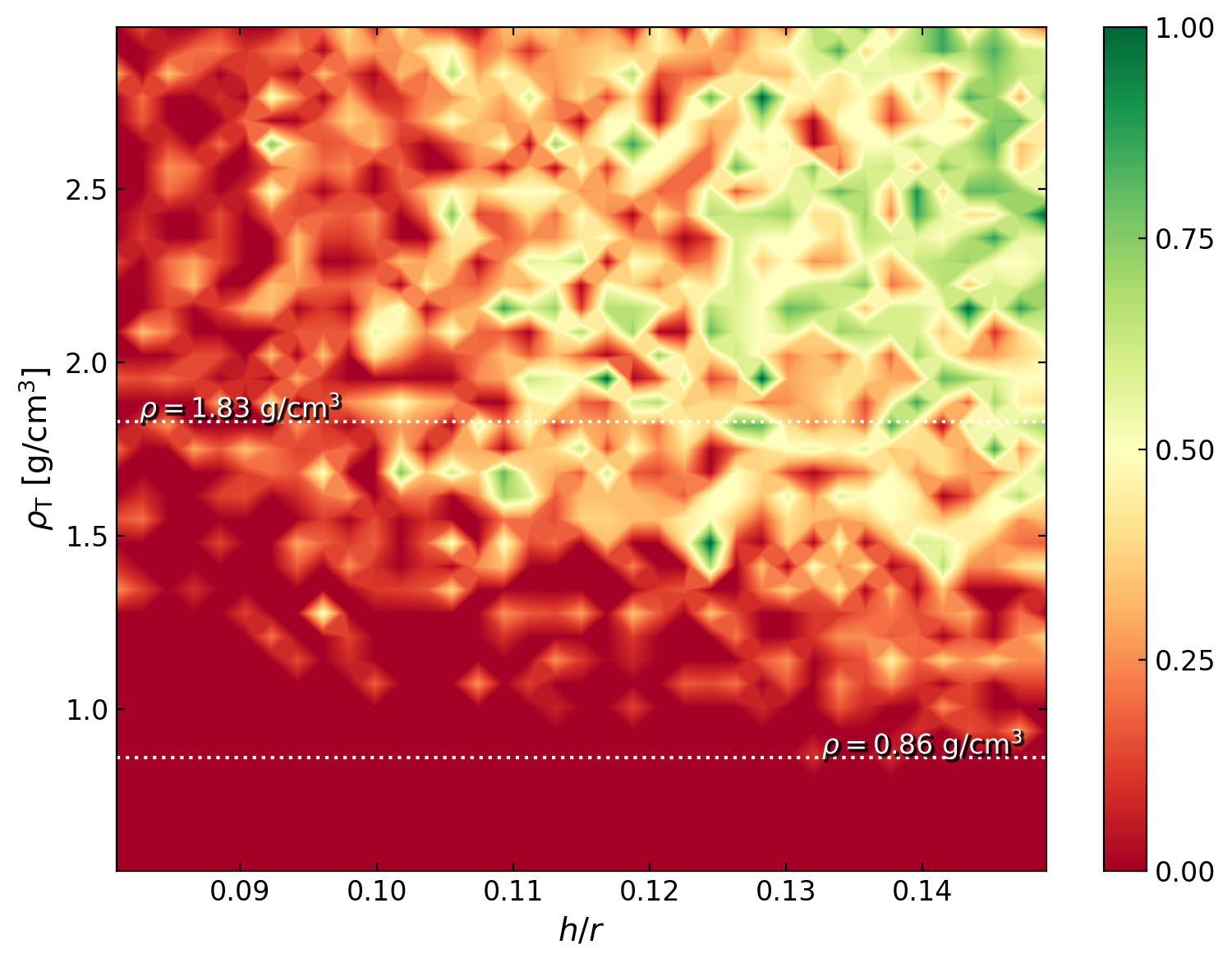}
    \caption{A grid of outcomes for 10,000 simulations as described in the text. Color corresponds to the success rate of Io's delivery of \textit{both} Amalthea and Thebe to the inner disk edge, \textit{while} replicating their proper modern ordering ($a_\text{A}<a_\text{T}$). Dotted density lines indicate Amalthea- and Callisto-like densities, corresponding to the exemplary runs shown in  Figures \ref{fig:TopTheMigPanels} and \ref{fig:BotTheMigPanels}}
    \label{fig:rhoThehMap}
\end{figure}

From the above simulations, we thus see that the critical material density of Thebe -- necessary for the ordered success of the resonant transport model -- clearly lies somewhere between the known densities of Amalthea and Callisto. To quantify this threshold further, we carried out 10,000 simulations, varying $\rho_\text{T}$, along with $h/r$. Figure \ref{fig:rhoThehMap} displays the outcomes of this suite of numerical experiments, with color indicating the success rate of Thebe and Amalthea reaching the magnetospheric cavity in the proper order of their modern semi-major axes. The grid readily illustrates the two criteria, with a relationship of success to both the disk-aspect ratio (governing the vigor of overstability), and the density of Thebe (determining the relative drag timescale between the two small moons). Most notably, setting an Amalthea-like density for Thebe yields a near-zero success rate for all realistic disk models, with the figure showing that Thebe's density proves to be no less than $\rho_{\text{T}}\sim 1.0$ g/cm$^3$ for success, and only becomes favorable to success at $\rho_{\text{T}} \gtrsim 1.4$ g/cm$^3$.

\section{Discussion}
\label{sect:Discussion}

While our previous work \citep{Brunton&Batygin25} placed a constraint on the \textit{primordial} nature of the Jovian system ($h/r\gtrsim0.08$), in this work we have derived a prediction of the scenario which could in principle be tested in the \textit{present-day} Jovian system. We submit that Thebe is overdense in comparison to its larger neighbor, Amalthea, with $\rho_\text{T} \gtrsim 1.0$ g/cm$^3 > \rho_\text{A}$ (for a mass of $m_\text{T}\gtrsim 5\times 10^{20}$ g), and likely closer to $\rho_\text{T} \approx 1.4$ g/cm$^3$. Values lower than our estimates would preclude the possibility for Thebe to have terminated at an orbit between Io and Amalthea upon delivery to its modern orbital neighborhood. That said, we acknowledge that our determination may be subject to stated assumptions in our disk model, and/or implicit assumptions about the dynamical history of Thebe and Amalthea. As such, there are, in principle, avenues in which our inference may be wrong, and we shall now examine such possibilities. 

\subsection{Disk flaring}
\label{flaring}

In our numerical experiments, we modeled the effects of the satellite-disk interactions while maintaining a constant aspect ratio with $\beta = 0$, as expected from viscous evolution in an active circumplanetary disk (see \citealt{Bitsch+15,Adams&Batygin25} and references therein). This expectation is not arbitrary, and is rooted in a self-consistent treatment of disk accretion in the active disk limit. But we should briefly justify this assumption since we have found it does indeed play a role in our final result.

The circumjovian disk is expected to be dominated by viscous heating, with irradiation from the young Jupiter providing only a minor correction (see e.g., \citealt{BatyginMorbidelli20}). In this limit where viscous heating dominates \textit{and} the disk is isothermal, the temperature profile scales with $r$ as $T \approx [(3\dot{M}\Omega^2/(16\,\pi\,\sigma))\,\sqrt{R_\text{H}/r}]^{1/4}$, with the accretion rate, $\dot{M}$, orbital frequency, $\Omega$, Stefan-Boltzmann constant, $\sigma$, and Hill radius, $R_\text{H}$.
It is then easy to show under this prescription, that the aspect ratio (proportional to the sound speed, $h/r\propto c_\text{s}=\sqrt{k_\text{b}T/\mu}$, with Boltzmann constant, $k_\text{b}$, and mean molecular weight, $\mu$) must scale as $h/r \propto r^{1/16}$. Moreover, in the limit of an optically thick disk (see e.g., \citealt{Batygin&Morbidelli22}), with dust-to-gas ratio, $f$, opacity, $\kappa$, and turbulence parameter, $\alpha$, the midplane temperature goes as $T\approx[3\,f\,\kappa\,\mu\,\dot{M}^2\,\Omega^3/(64\,\pi^2\,k_\text{b}\,\sigma\,\alpha)]^{1/5}$, which translates to an even slighter dependence on $r$ for the aspect ratio, with $h/r\propto r^{1/20}$. Ergo, with $\beta\sim1/20$--$1/16$, approximating $h/r$ as constant holds true given our stated assumptions. Nonetheless, since other models of the early Jovian system envision a more passive nebula (e.g., \citealt{RonnetJohansen20}), we will examine the alternative scenario for completeness. 

\begin{figure}[b]
    \centering
    \includegraphics[width=1\linewidth]{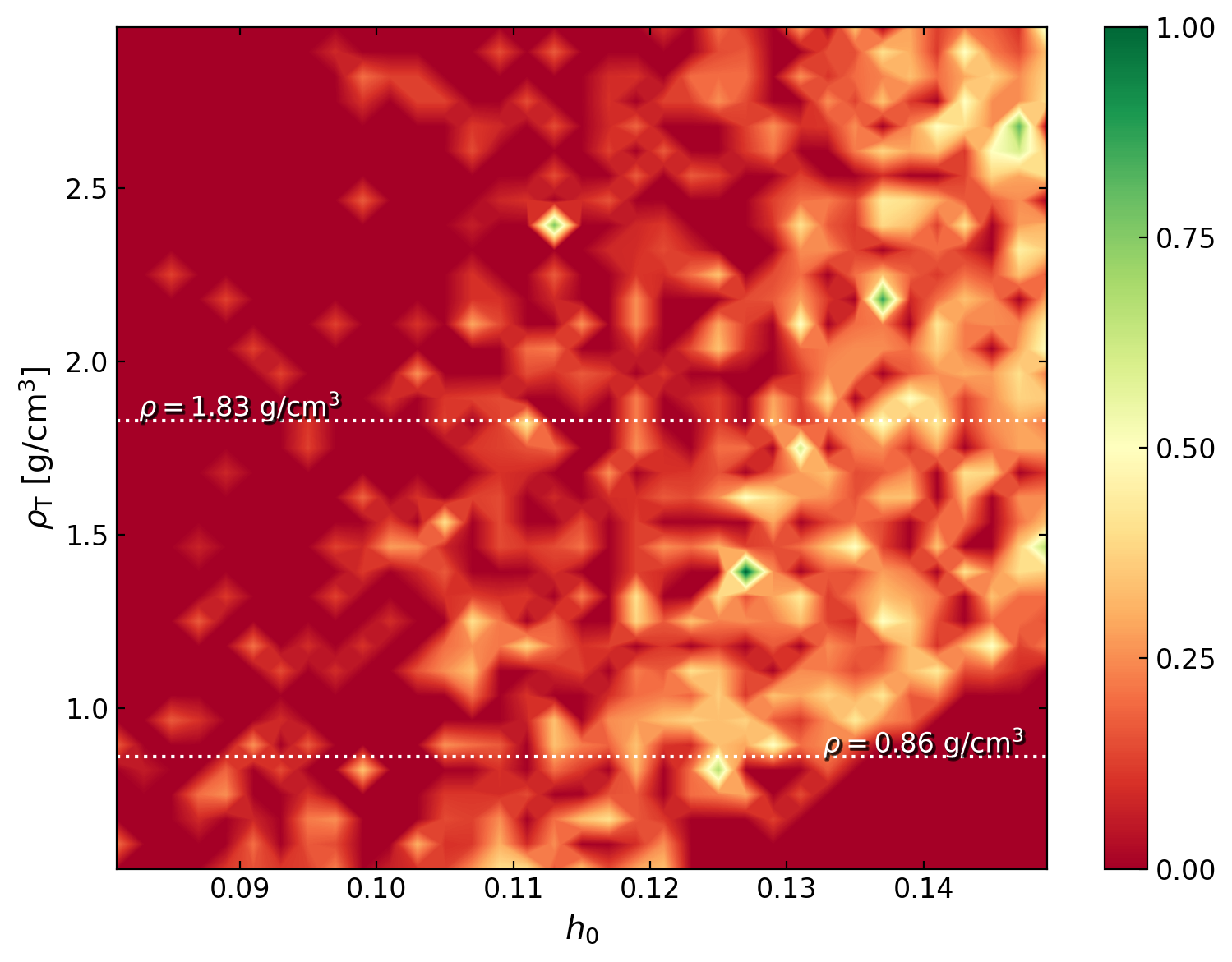}
    \caption{A similar grid of outcomes for 10,000 simulations as that of Figure \ref{fig:rhoThehMap}, now instead with minor flaring in the disk, as $\beta=1/4$, and varying the reference aspect ratio between $h_0=$ 0.08--0.15. Since $h/r = h_0(r/r_0)^\beta$, this adjustment \textit{decreases} the overall success rate throughout parameter space, with a slight but notable exception between $h_0\approx0.09$--0.12 and $\rho_{T}\lesssim1.0$ g/cm$^3$, in which the success rate \textit{increases}.}
    \label{fig:BetarhoThehMap}
\end{figure}

If we were to slightly relax the above conditions, imagining instead a non-negligible contribution of Jupiter's irradiation to the thermal energy balance of the disk, we would expect minor flaring in the nebula's vertical structure. Generically, the two criteria outlined in \S \ref{sec:intro} for replicating the orbital architecture would still apply. However, this adjustment creates a small exception to the necessary result that Thebe's \textit{density} must be higher. Instead, there is a narrow window of parameter space in which a density of $\rho_\text{T} \lesssim \rho_\text{A}$ is amenable in the flared-disk model. Figure \ref{fig:BetarhoThehMap} illustrates this point, which is a heat map of outcomes for 10,000 simulations with the same setup described for Figure \ref{fig:rhoThehMap}, albeit now with $\beta = 1/4$. This entails that in these runs, we are specifically varying the reference aspect ratio, $h_0$, since the disk aspect ratio takes on the typical power law profile, $h/r=h_0(r/r_0)^\beta$. It is clear from the figure that a flared disk actually \textit{diminishes} the overall success rate of the resonant transport mechanism -- a conclusion already implied in our previous results since the increase in $\beta$ is essentially equivalent to lowering the relative $h/r$. Nevertheless, in the specific range of $h_0\approx0.09$--0.12, and $\rho_{T} \lesssim \rho_{A}$, the success rate is actually \textit{enhanced} relative to the purely viscous case. Therefore, a measured \textit{under-dense} Thebe -- in contrast to our expectation above -- would have interesting ramifications for our theoretical model(s) of Jupiter's primordial environment. But in light of the fact that the disk must accommodate a considerable accretion rate, $\dot{M}$, and be relatively hot (to accommodate $h/r\gtrsim0.08$), we expect it to be largely viscosity driven, and thus do not expect it to be flared.

\subsection{Disruption and re-coagulation}

We are implicitly working under the assumption that the modern Amalthea and Thebe maintain characteristics representative of their disk-epoch conglomeration. But perhaps one or both of the moons experienced catastrophic disruption and re-coagulation, such that the current radii and densities do not reflect their primordial values? This is certainly a possibility given that gravitational focusing by Jupiter can enhance the relative rate of catastrophic disruption for its small inner moons \citep{Nesvorny+23}. In fact, we expect this exact scenario may very well be the case for Metis and Adrastea, which are far smaller in radii than their inner moon counterparts, and are separated from each other by only $\Delta a \sim 1000$ km -- well interior of the Roche limit. While difficult to (dis)prove definitively, such a scenario may render all inferences on the physical structure of the disk or the individual properties of the satellites unsettled. Nevertheless, this would not refute the resonant transport theory itself.

\subsection{Conclusion}

All considered, we still put forth the most likely outcome implied by our numerical experiments: Thebe possesses a density of $\rho_\text{T} \gtrsim 1.0$ g/cm$^3$. Our prediction also aligns well with a conclusion reached by \citet{Cuk+23} -- the only other study to make a deliberate inference about Thebe's density -- who suggested that Thebe may be more dense than Amalthea in order to remain immune from its own demise via ``sesquinary catastrophe."\footnote{However, \citet{Cuk+23} also make it clear that this could instead be compensated by Thebe's internal material strength rather than a larger density.} 

Overall, our work offers an observational avenue towards falsification or confirmation of our proposed model to explain the architecture of Jupiter's regular moons. A flyby of Thebe with the \textit{Juno} spacecraft could extract gravitational data to determine Thebe's mass and density (as was done with the \textit{Galileo} probe for Amalthea by \citealt{Anderson+02}). Moreover, JWST NIRSpec of Thebe's surface could provide immediate resolution on its ice composition, while also pointing to the location of its origin in the circumjovian disk.  

\begin{acknowledgments}
\textit{Acknowledgments}.
IB is grateful to the Ahmanson Foundation for financial support, and KB is thankful for the support of the David \& Lucile Packard Foundation, along with the National Science Foundation (grant number: AST 2408867). Both authors thank Caltech and $^3$CPE. 
\end{acknowledgments}

\bibliographystyle{aasjournal}
\bibliography{Thebe}

\end{document}